\documentclass[%
 reprint,
 amsmath,amssymb,
 aps,
]{revtex4-2}

\usepackage{graphicx}
\usepackage{dcolumn}
\usepackage{bm}
\usepackage{hyperref}

\usepackage{amsmath,amssymb,amsthm,mathtools}
\usepackage{graphicx}
\usepackage{xcolor}
\usepackage{physics}
\usepackage{enumitem}

\newcommand{\ii}{\mathrm{i}}
\newcommand{\ee}{\mathrm{e}}

\newcommand{\rb}{\mathbf{r}}
\newcommand{\qb}{\mathbf{q}}
\newcommand{\kb}{\mathbf{k}}
\newcommand{\Rb}{\mathbf{R}}
\newcommand{\Sb}{\mathbf{S}}

\newcommand{\chiphys}{\chi_{\mathrm{phys}}}
\newcommand{\chiMF}{\chi_{0}}
\newcommand{\Heff}{H_{\mathrm{eff}}}
\newcommand{\HLS}{H_{\mathrm{LS}}}
\newcommand{\Hbus}{H_{\mathrm{bus}}}
\newcommand{\Dcal}{\mathcal{D}}
\newcommand{\Lcal}{\mathcal{L}}
\newcommand{\Qbus}{Q_{\mathrm{bus}}}

\newcommand{\Dsl}{\ensuremath{\mathrm{DSL}}}

\newcommand{\physop}{\mathcal{O}_{\mathrm{spin}}}

\begin{document}

\preprint{APS/123-QED}

\title{Manuscript title}

\title{Coherent Exchange and Decoherence in Dirac-Spin-Liquid Quantum Interconnects}

\author{Dibakar Yadav}
\author{Rana Pratap}%
\affiliation{%
Department of Electrical Engineering,  Indian Institute of Technology Madras, Chennai 600036, India
}%

\date{\today}

\begin{abstract}
We develop a susceptibility-based open-system theory for two localized qubits coupled through a candidate two-dimensional $\mathrm{U}(1)$ Dirac-spin-liquid-like bath. The central input is the gauge-invariant retarded physical spin susceptibility $\chiphys^R(\qb,\omega)$ of the bath. We show that this single response kernel controls both coherent and dissipative qubit dynamics: its real part generates the nonlocal mediated exchange, while its absorptive part determines relaxation and dephasing through the equilibrium noise spectrum. This gives a unified reduced two-qubit description in which the usefulness of the bath as an entanglement bus is governed by the competition between susceptibility-mediated exchange and bath-induced decoherence. As an analytically transparent benchmark, we evaluate the spinon mean-field Dirac susceptibility and recover the static algebraic exchange $J_{\mathrm{eff}}(R)\propto J_{\rm local}^2/(v_F R^3)$, together with pseudogap-suppressed relaxation $\Gamma_1\propto J_{\rm local}^2\omega_0^3/v_F^4$. We then formulate a beyond-mean-field extension in which gauge-field dressing and other interaction effects are absorbed into a dressed physical susceptibility, without changing the reduced qubit-sector mapping. The resulting framework provides a direct route from the many-body spin response of a correlated two-dimensional bath to reduced-dynamics simulations of entanglement generation, coherence loss, and the operational phase space of a candidate Dirac spin-liquid quantum interconnect.
\end{abstract}

\maketitle


\section{\label{sec:level1}Introduction}

A central problem in solid-state quantum information processing is to realize a nonlocal quantum interconnect capable of entangling spatially separated qubits~\cite{LossDiVincenzo1998,Cirac1997,Bose2003,Christandl2004}. At the same time, such an interconnect must not overwhelm the qubit sector with the very noise it introduces. In the present work, the candidate interconnect is a strongly correlated spin bath, motivated by the broader physics of quantum spin liquids and emergent gauge-field matter~\cite{SB2017,Broholm2020}. We assume that, over the relevant frequency and distance scales, its response is governed by a two-dimensional $\mathrm{U(1)}$ Dirac-spin-liquid-like bath susceptibility~\cite{Hermele2004,Hermele2005,Ran2007,Hermele2008,Song2020PRX}. The central question is whether such a bath can mediate useful exchange before bath-induced decoherence dominates~\cite{Legg2019RKKY}.

This viewpoint suggests a relatively unexplored spin-bus application of Dirac-spin-liquid(DSL) physics. A correlated two-dimensional spin medium can mediate coupling through its nonlocal spin response, while its planar geometry is naturally compatible with van der Waals or other atomically thin heterostructures that host qubits or quantum emitters ~\cite{Onizhuk2021,Tran2016}. The relevant figure of merit is therefore not exchange alone, but the balance between susceptibility-mediated exchange and the decoherence generated by the same bath.

The appropriate framework for this problem is linear response. In close analogy with dielectric response or charge-density response in electronic matter, the bath is characterized by a causal retarded susceptibility~\cite{Kubo1966,Mahan2000}. Here the relevant object is the gauge-invariant physical spin susceptibility $\chiphys^R(\qb,\omega)$. Once this kernel is specified, the reduced two-qubit dynamics follows systematically. The dispersive part of the susceptibility generates coherent exchange and local Lamb shifts, while the absorptive part fixes the dissipative channels through the corresponding equilibrium noise spectrum~\cite{Kubo1966,Mahan2000,Clerk2010RMP,BP2002book}.In this role, the bath susceptibility acts as a nonlocal, frequency-dependent spin-transfer function for the qubits.

At this stage, it is important to identify the bath operator that the qubits actually probe. A $\mathrm{U(1)}$ Dirac spin liquid is not simply a gas of free Dirac spinons; its low-energy description is an interacting gauge theory of massless Dirac spinons coupled to an emergent dynamical $\mathrm{U(1)}$ gauge field~\cite{Hermele2004,Hermele2005,Ran2007,Song2020PRX}. In the standard spin-$1/2$ convention used below, the reference theory contains four two-component Dirac spinon flavors~\cite{Ran2007,Song2020PRX}. The qubits therefore couple not to the gauge-charged spinon fields themselves, but to gauge-invariant physical spin operators of the bath~\cite{Wen2002,WenBook2004,Hermele2005,SB2017}. This distinction is conceptually important, but it is technically clean in the present framework. It does not modify the structure of the reduced open-system derivation; it changes only the physical susceptibility that must be supplied as bath input. We exploit this separation throughout: the reduced qubit-sector mapping is formulated entirely in terms of the physical susceptibility, while the specific bath physics enters only through the choice of that susceptibility. 

\begin{figure}
    
    \includegraphics[width= 0.94\columnwidth]{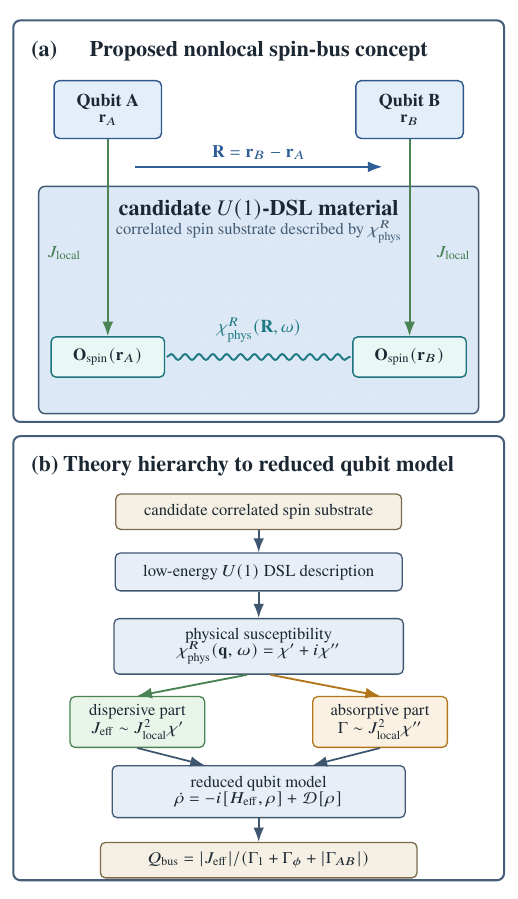}
    \caption{Device concept and theory hierarchy for the susceptibility-based spin-bus framework.
(a) Two localized qubits couple locally to gauge-invariant spin operators of a candidate $\mathrm{U(1)}$-DSL material; the nonlocal physical spin susceptibility $\chi_{\rm phys}^{R}(\mathbf R,\omega)$ mediates the bath response between the two positions.
(b) The correlated substrate is represented by its physical susceptibility, whose dispersive and absorptive parts generate the coherent exchange and dissipative rates entering the reduced two-qubit master equation.}
    \label{fig:device_theory_flow}
\end{figure}

With this separation in hand, we treat the bath within linear-response theory and the qubit dynamics within weak-coupling open-system theory~\cite{Davies1974,BP2002book}. The bath-specific input is the choice of a Dirac-spin-liquid-like susceptibility, which may be taken either as the analytically transparent mean-field benchmark or as a dressed susceptibility incorporating interaction effects at a phenomenological level.  Figure~\ref{fig:device_theory_flow} summarizes the device-level picture and the theoretical reduction used throughout the paper. Panel~(a) shows the proposed nonlocal spin-bus geometry: two localized qubits couple locally to gauge-invariant spin operators of a candidate $\mathrm{U(1)}$-DSL material, while the nonlocal bath response propagates spin correlations between the two positions. Panel~(b) gives the corresponding theory hierarchy. The correlated substrate is represented by its physical spin susceptibility, whose dispersive and absorptive parts generate, respectively, the coherent exchange and the dissipative rates entering the reduced two-qubit master equation. The schematic therefore presents a susceptibility-based framework for a candidate many-body interconnect rather than a material-specific device blueprint.

The paper is organized around this logic. Section~\ref{sec:low_energy} introduces the qubit sector, the low-energy gauge-theory bath description, and the gauge-invariant susceptibility that serves as the central bath input. Section~\ref{sec:reduced_dynamics} derives the reduced two-qubit master equation in the weak-coupling regime and shows how the same bath kernel generates both the coherent bus Hamiltonian and the dissipator. Section~\ref{sec:mean_field} develops an analytically transparent neutral Dirac-spinon mean-field benchmark, recovering the algebraic exchange $J_{\mathrm{eff}}(R)\propto J_{\rm local}^2/(v_F R^3)$ together with pseudogap-suppressed relaxation. Section~\ref{sec:beyond_mean} extends the analysis to a dressed physical susceptibility without altering the reduced qubit-sector derivation. Section~\ref{sec:sim} presents the reduced-dynamics simulations and operational observables, while Section~\ref{sec:diss} summarizes the main physical conclusions, their limitations, and the operating-window interpretation. Technical details are collected in the Appendices. Throughout, the emphasis is on the conditional viability of a candidate many-body bath as a spin bus, not on assuming that a specific material realization is already an established and asymptotically stable $\mathrm{U(1)}$ Dirac spin liquid.

\section{Low-energy theory of the $\mathrm{U(1)}$ Dirac spin-liquid bath}
\label{sec:low_energy}

To formulate the qubit--bath problem at low energies, we separate the theory into a localized qubit sector, a critical DSL bath sector, and the gauge-invariant bath operator through which the qubits couple to that bath. The main structural point of this section is that, once this operator is identified, the entire spin-bus problem reduces to the corresponding physical spin susceptibility of the bath.

\subsection{Qubit sector}

We consider two localized spin-$1/2$ qubits, labeled $A$ and $B$, located at
\begin{equation}
\rb_A = 0, \qquad \rb_B = \Rb.
\label{eq:qubit_positions}
\end{equation}
Their bare Hamiltonian is
\begin{equation}
H_q = \frac{\omega_0}{2}\sigma_A^z + \frac{\omega_0}{2}\sigma_B^z,
\label{eq:Hq}
\end{equation}
where $\omega_0$ is the qubit splitting and $\sigma_m^a$ are Pauli matrices acting on qubit $m\in\{A,B\}$. Throughout the numerical analysis we use dimensionless units with \(\hbar=1\) and \(k_B=1\), so that energies, frequencies, and temperatures are measured in the same units. It is convenient to define spin operators
\begin{equation}
S_m^a = \frac{1}{2}\sigma_m^a,
\label{eq:spin_ops}
\end{equation}
so that Eq.~\eqref{eq:Hq} may equivalently be written as
\begin{equation}
H_q = \omega_0 S_A^z + \omega_0 S_B^z.
\label{eq:Hq_spin}
\end{equation}
This form makes the later frequency decomposition transparent: \(S_m^z\) preserves the qubit energy, while \(S_m^\pm\) generate transitions at frequencies \(\pm\omega_0\).

\subsection{Gauge-theory description of the bath}

The reference $\mathrm{U(1)}$ Dirac spin liquid is described at low energies by an interacting $2+1$-dimensional gauge theory of massless Dirac spinons coupled to an emergent compact $\mathrm{U(1)}$ gauge field~\cite{Hermele2004,Hermele2005,Ran2007,Hermele2008,Song2020PRX}. In the standard spin-$1/2$ DSL convention used for common two-dimensional lattice realizations, the continuum theory contains four two-component Dirac spinon flavors~\cite{Ran2007,Song2020PRX}. In Euclidean notation we write the bath Lagrangian as
\begin{equation}
\Lcal_{\Dsl}^{E}
=
\sum_{\alpha=1}^{4}
\bar\psi_\alpha
\gamma^\mu
(\partial_\mu-\ii a_\mu)
\psi_\alpha
+
\frac{1}{4e^2} f_{\mu\nu} f_{\mu\nu}
+ \cdots ,
\label{eq:L_dsl}
\end{equation}
where $\mu,\nu\in\{\tau,x,y\}$ are Euclidean spacetime indices, $\psi_\alpha$ are emergent Dirac spinons, and $a_\mu$ is the emergent dynamical $\mathrm{U(1)}$ gauge field. The tensor
$f_{\mu\nu}=\partial_\mu a_\nu-\partial_\nu a_\mu$ is the corresponding gauge-field strength, so that $f_{\mu\nu}f_{\mu\nu}/(4e^2)$ is the Maxwell-like stiffness term for the emergent gauge field. We use $\bar\psi_\alpha$ to denote the Dirac conjugate spinon field, conventionally written as $\bar\psi_\alpha=\psi_\alpha^\dagger\gamma^0$ in a chosen gamma-matrix convention. Roman indices $a,b\in\{x,y,z\}$ are reserved for physical spin components. The ellipsis denotes symmetry-allowed short-distance terms whose detailed form is not needed for the susceptibility-based reduction used below.

Equation~\eqref{eq:L_dsl} serves as a reference low-energy field theory for a candidate DSL bath. The compactness of the emergent gauge field and the symmetry quantum numbers of monopole operators are important for the ultimate infrared stability of a particular lattice realization~\cite{Hermele2004,Song2020PRX}. That stability question is not the focus here. Instead, we take a conditional viewpoint and ask what reduced qubit dynamics follow if the bath exhibits DSL-like physical spin response over the frequency and length scales relevant to the qubits. This viewpoint also fixes the appropriate bath operator: since spinons are gauge charged and are not themselves physical observables, external spin probes must couple to gauge-invariant physical spin operators of the bath~\cite{Wen2002,WenBook2004,Hermele2005,SB2017}.

We therefore model the local qubit--bath coupling by the minimal exchange form
\begin{equation}
H_{\mathrm{int}}
=
J_{\rm local}\,\Sb_A\cdot \bm{\physop}(\rb_A)
+
J_{\rm local}\,\Sb_B\cdot \bm{\physop}(\rb_B),
\label{eq:Hint_general}
\end{equation}
where $J_{\rm local}$ is the local qubit--bath exchange coupling and
$\bm{\physop}=(\physop^x,\physop^y,\physop^z)$ is the vector of gauge-invariant physical spin operators of the bath. At the reduced-qubit level, this coupling generates a susceptibility-mediated indirect interaction, analogous in structure to spin-liquid-mediated RKKY-type exchange~\cite{Legg2019RKKY}, but applied here to localized qubits and treated within an open-system framework.

The microscopic content of $\physop^a$ depends on the lattice realization and on how the physical spin operator projects into the low-energy theory. It may include long-wavelength fermion bilinears and, in some cases, contributions from other gauge-invariant critical operators such as monopole operators~\cite{Hermele2005,Hermele2008,Song2019NatComm,Song2020PRX}. The corresponding projection factors, flavor-counting factors, and operator-normalization constants are absorbed into the overall normalization of the physical susceptibility $\chi_{\mathrm{phys}}^R$. Thus the reduced qubit-sector formalism developed below does not require specifying the microscopic decomposition of the bath operator $\physop^a$; it requires only its gauge-invariant susceptibility.

\subsection{Physical susceptibility of the bath}

The central bath quantity is therefore the retarded gauge-invariant spin susceptibility
\begin{equation}
\begin{aligned}
\chi_{\mathrm{phys},ab}^R(\rb-\rb',t)
&=
-\ii\theta(t)\,
\bigl\langle
\bigl[
\physop^a(\rb,t),\,
\\
&\hspace{1.2em}\physop^b(\rb',0)
\bigr]
\bigr\rangle_{\Dsl}.
\label{eq:chi_phys_realspace}
\end{aligned}
\end{equation}

Here $\theta(t)$ is the Heaviside step function, which enforces causality in the retarded response. Equation~\eqref{eq:chi_phys_realspace} is the spin analog of a retarded density-response function in electronic linear-response theory~\cite{Kubo1966,Mahan2000}. With the source convention
\begin{equation}
H_{\mathrm{src}}(t)
=
\sum_b
\int \dd^2r\;
h_b(\rb,t)\physop^b(\rb,t),
\label{eq:source_coupling}
\end{equation}
linear response gives
\begin{equation}
\begin{aligned}
\delta\langle \physop^a(\rb,t)\rangle
&=
\sum_b
\int dt' \int \dd^2r'\;
\\
&\quad\times
\chi_{\mathrm{phys},ab}^R(\rb-\rb',t-t')\,h_b(\rb',t').
\end{aligned}
\label{eq:linear_response_physop}
\end{equation}
Thus $\chi_{\mathrm{phys},ab}^R$ is the causal spin-response kernel of the bath.

For a homogeneous and stationary bath, the response depends only on coordinate and time differences, so it is natural to work in momentum--frequency space. The Fourier transform of the real-space retarded susceptibility is
\begin{equation}
\begin{aligned}
\chi_{\mathrm{phys},ab}^R(\qb,\omega)
&= \int \dd t \int \dd^2r\;
   \ee^{\ii(\omega t-\qb\cdot\rb)}
   \chi_{\mathrm{phys},ab}^R(\rb,t) \\
&= \chi_{\mathrm{phys},ab}'(\qb,\omega)
   + \ii\,\chi_{\mathrm{phys},ab}''(\qb,\omega),
\end{aligned}
\label{eq:chi_phys_fourier}
\end{equation}
where the real part $\chi_{\mathrm{phys},ab}'$ controls dispersive energy shifts and coherent exchange, whereas the imaginary part $\chi_{\mathrm{phys},ab}''$ controls absorption and, through the associated equilibrium noise spectrum, the dissipative channels of the reduced qubit dynamics~\cite{Kubo1966,Mahan2000,Clerk2010RMP}. In practical calculations one often evaluates the corresponding Matsubara correlator in the Euclidean theory and then analytically continues $\ii\Omega_n\to \omega+\ii0^+$ to obtain Eq.~\eqref{eq:chi_phys_fourier}~\cite{Mahan2000}.

The equilibrium fluctuation spectra associated with the same operator are fixed by the absorptive response through the fluctuation-dissipation theorem~\cite{Kubo1966,Mahan2000}. It is useful, however, to distinguish two related objects. The symmetrized spectrum measures total equilibrium fluctuation power, whereas finite-frequency transition rates in the reduced master equation are controlled by the ordered, or unsymmetrized, bath spectrum~\cite{Clerk2010RMP,GardinerZoller,BP2002book}. This distinction is made explicit in Sec.~\ref{sec:reduced_dynamics}. The susceptibility-based formulation is nevertheless the same: once $\chi_{\mathrm{phys}}^R$ is specified, both the coherent bath-mediated bus interaction and the dissipative qubit dynamics are determined by gauge-invariant two-point functions of $\physop^a$.

\section{Reduced two-qubit dynamics from the physical susceptibility} \label{sec:reduced_dynamics}

We now translate the susceptibility-based bath description of Sec.~\ref{sec:low_energy} into a reduced dynamical theory for the qubits. The central point is that, in the weak-coupling regime appropriate to a mediated spin bus, the bath affects the qubits only through the same gauge-invariant response kernel introduced above, now reorganized into coherent and dissipative sectors of the reduced evolution.

The total Hamiltonian is
\begin{equation}
H = H_q + H_{\Dsl} + H_{\mathrm{int}}.
\label{eq:Htotal}
\end{equation}
Our goal is to eliminate the bath and obtain an effective description of the two qubits alone. In the weak-coupling regime appropriate to a mediated bus, the bath remains close to equilibrium and the leading nontrivial qubit-qubit coupling is second order in the local qubit--bath coupling $J_{\rm local}$: one interaction event injects a disturbance into the bath, and a second interaction event samples the propagated response.

\subsection{Second-order reduced dynamics}

In the interaction picture with respect to $H_q+H_{\Dsl}$, and within the standard Born--Markov weak-coupling approximation, the reduced density matrix of the qubits obeys, to second order in the local qubit--bath coupling $J_{\rm local}$~\cite{Redfield1957,Davies1974,BP2002book},
\begin{equation}
\begin{aligned}
\dot\rho(t)
&=
-\int_0^{\infty} \dd\tau\; \Tr_{\Dsl}
\Big[
H_{\mathrm{int}}(t),
\\
&\qquad\big[
H_{\mathrm{int}}(t-\tau),\rho(t)\otimes \rho_{\Dsl}
\big]
\Big].
\end{aligned}
\label{eq:born_markov_start}
\end{equation}
where $\rho_{\Dsl}$ is the equilibrium bath state. Equation~\eqref{eq:born_markov_start} is the standard Born--Markov memory kernel. The reduced qubit dynamics is generated by two interaction events separated by a bath memory time $\tau$, after which the bath is traced out. In the present context, Eq.~\eqref{eq:born_markov_start} is the open-system analog of integrating out a reservoir to obtain a second-order self-energy for the qubit sector.

For the local exchange coupling in Eq.~\eqref{eq:Hint_general}, the interaction-picture Hamiltonian entering Eq.~\eqref{eq:born_markov_start} is
\begin{equation}
H_{\mathrm{int}}(t)
=
J_{\rm local}
\sum_{m=A,B}
\sum_{a=x,y,z}
S_m^a(t)\,\physop^a(\rb_m,t).
\label{eq:Hint_t}
\end{equation}
This form makes it clear that each interaction vertex contributes one qubit operator and one bath operator. Since Eq.~\eqref{eq:born_markov_start} is second order in $H_{\mathrm{int}}$, the bath trace contains two bath operators and therefore generates the two-point correlation functions
\begin{equation}
C_{mn}^{ab}(\tau)
=
\ev{\physop^a(\rb_m,\tau)\,\physop^b(\rb_n,0)}_{\Dsl}.
\label{eq:bath_corr_matrix}
\end{equation}
Equivalently, with the source convention of Eq.~\eqref{eq:source_coupling}, the qubit spins act as localized source fields for the bath operator. Thus all coherent and dissipative effects induced by the DSL enter the reduced qubit dynamics through the correlation matrix $C_{mn}^{ab}(\tau)$. This is the reduced-dynamics counterpart of the response-function viewpoint of Sec.~II: the bath enters only through gauge-invariant two-point functions, rather than through bare microscopic spinon fields.

To identify which frequency components of the bath are probed by the qubits, we decompose the qubit operators into parts that evolve with definite Bohr frequencies under $H_q$, as in the standard frequency-resolved construction of weak-coupling master equations~\cite{Davies1974,BP2002book},
\begin{equation}
S_m^a(t) = \sum_{\omega} \ee^{-\ii\omega t} S_m^a(\omega),
\label{eq:bohr_decomp}
\end{equation}
where $\omega$ runs over the Bohr frequencies of the qubit sector. For Eq.~\eqref{eq:Hq_spin}, the nonzero components are
\begin{equation}
S_m^z(0)=S_m^z;\;
S_m^+(-\omega_0)=S_m^+;\;
S_m^-(+\omega_0)=S_m^-.
\label{eq:bohr_components}
\end{equation}
Accordingly, the bath is sampled only at $\omega\in\{0,\pm\omega_0\}$, corresponding respectively to longitudinal fluctuations and spin-flip processes. This decomposition is useful because it makes explicit which parts of the bath spectrum control dephasing and which control qubit transitions.

Substituting Eq.~\eqref{eq:bohr_decomp} into Eq.~\eqref{eq:born_markov_start} and performing the $\tau$ integration defines the one-sided Fourier transforms
\begin{equation}
\begin{aligned}
\mathcal{K}_{mn}^{ab}(\omega)
&=
J_{\rm local}^2 \int_0^\infty \dd\tau\,
\ee^{\ii\omega\tau} C_{mn}^{ab}(\tau)
\\
&=
\frac{1}{2}\gamma_{mn}^{ab}(\omega)
+\ii\Lambda_{mn}^{ab}(\omega),
\end{aligned}
\label{eq:Gamma_def}
\end{equation}
which are the complex, frequency-resolved Born--Markov kernels sampled by the qubits. More precisely, $\gamma_{mn}^{ab}(\omega)$ is the dissipative Hermitian rate matrix, while $\Lambda_{mn}^{ab}(\omega)$ parameterizes the principal-value part that generates coherent Lamb-shift and exchange terms. The full derivation, including the intermediate Redfield form and the organization of the secular frequency sectors, is given in Appendix~\ref{app:redfield}. Here it suffices to note that the secular approximation discards terms oscillating as $\ee^{\ii(\omega'-\omega)t}$ with $\omega\neq \omega'$, leaving a reduced equation that separates naturally into coherent and dissipative contributions on the slow qubit timescale.

After the standard Born--Markov--secular reduction, the master equation assumes the Gorini--Kossakowski--Sudarshan--Lindblad form~\cite{Gorini1976,Lindblad1976,Davies1974,BP2002book}
\begin{equation}
\dot\rho
=
-\ii[H_q+\HLS+\Hbus,\rho]
+
\Dcal[\rho].
\label{eq:generic_master}
\end{equation}
The term $\HLS$ contains local Lamb shifts, $\Hbus$ contains the nonlocal coherent interaction generated by the bath, and $\Dcal$ is the dissipator. Equation~\eqref{eq:generic_master} is therefore the reduced qubit-sector form of the statement already made at the response-function level: the same bath kernel controls both coherent exchange and decoherence.

\subsection{Coherent bus Hamiltonian}

The coherent bath-induced contribution is encoded in the dispersive part $\Lambda_{mn}^{ab}(\omega)$, which is related to the dispersive real part of the physical susceptibility. This is the same susceptibility-mediated logic underlying indirect-exchange constructions such as RKKY interactions, here applied to a spin-liquid bath and projected onto the qubit sector~\cite{Ruderman1954,Kasuya1956,Yosida1957,Legg2019RKKY}. Separating the local and nonlocal pieces, one may write the nonlocal coherent interaction as
\begin{equation}
\begin{aligned}
\Hbus
&=
\sum_{a,b}
J_{\mathrm{eff}}^{ab}(\Rb,\omega_0)
S_A^a S_B^b,
\\
J_{\mathrm{eff}}^{ab}(\Rb,\omega_0)
&\propto
J_{\rm local}^2
\chi_{\mathrm{phys},ab}'(\Rb,\omega_0).
\end{aligned}
\label{eq:Hbus_general}
\end{equation}
This is the operator form of the transfer-function picture discussed above. Qubit $A$ drives the bath, the bath response propagates from $\rb_A$ to $\rb_B$, and qubit $B$ samples the propagated field. The bus Hamiltonian is therefore the coherent part of the same response kernel introduced in Sec.~II, now projected onto the qubit sector. In the static adiabatic limit,
\begin{equation}
\Hbus^{\mathrm{static}}
=
\sum_{a,b} J_{\rm local}^2 \chi_{\mathrm{phys},ab}'(\Rb,0) S_A^a S_B^b.
\label{eq:Hbus_static_general}
\end{equation}
If the bath is effectively SU(2)-symmetric at long wavelengths,
\begin{equation}
\chi_{\mathrm{phys},ab}^R(\qb,\omega)=\delta_{ab}\,\chiphys^R(\qb,\omega),
\label{eq:isotropic_general}
\end{equation}
and the real-space response is radial, \(\chiphys'(\Rb,0)=\chiphys'(R,0)\) with $R=|\Rb|$ Equation~\eqref{eq:Hbus_static_general} then reduces to the Heisenberg form
\begin{equation}
\begin{aligned}
\Hbus^{\mathrm{static}}
&=
J_{\mathrm{eff}}(R)\,\Sb_A\!\cdot\!\Sb_B,
\\
J_{\mathrm{eff}}(R)
&=
J_{\rm local}^2\chiphys'(R,0).
\end{aligned}
\label{eq:Hbus_heisenberg_general}
\end{equation}
The sign of $J_{\mathrm{eff}}$ is determined by the sign of the static susceptibility and therefore encodes whether the induced interaction is effectively ferro- or antiferromagnetic. 

\subsection{Dissipation and decoherence}

The dissipator is controlled by the same bath kernel through its absorptive part, but the object that enters a finite-frequency transition rate is the ordered, or unsymmetrized, bath spectrum rather than the symmetrized noise alone~\cite{Clerk2010RMP,GardinerZoller,BP2002book}. For the bath operators coupled to qubits $m$ and $n$, define the ordered bath spectrum
\begin{equation}
\begin{aligned}
\mathcal S_{mn}^{ab,>}(\omega)
&=
\int_{-\infty}^{\infty}\dd t\;\ee^{\ii\omega t}
\\
&\quad\times
\ev{\physop^a(\rb_m,t)\physop^b(\rb_n,0)}_{\Dsl}.
\end{aligned}
\label{eq:unsym_noise_spectrum}
\end{equation}
Here \(\mathcal S_{mn}^{ab,>}(\omega)\) denotes a bath noise spectrum and should not be confused with the qubit spin operator \(S_m^a\). In the weak-coupling Markovian limit, the rate matrix at Bohr frequency $\omega$ is~\cite{Redfield1957,Davies1974,BP2002book}
\begin{equation}
\gamma_{mn}^{ab}(\omega)
=
J_{\rm local}^2 \mathcal S_{mn}^{ab,>}(\omega).
\label{eq:rates_unsym}
\end{equation}
At equilibrium, detailed balance relates positive- and negative-frequency spectra, and the fluctuation-dissipation theorem gives~\cite{Kubo1966,Mahan2000,Clerk2010RMP}
\begin{equation}
\mathcal S_{mn}^{ab,>}(\omega)
=
\frac{2}{1-\ee^{-\beta\omega}}\,
\chi_{\mathrm{phys},ab}''(\rb_m-\rb_n,\omega),
\label{eq:unsym_fdt}
\end{equation}
up to the overall Fourier-transform and retarded-sign convention. Equivalently, if one defines the absorptive part to be positive for positive-frequency absorption, Eq.~\eqref{eq:unsym_fdt} should be read with that positive absorptive spectral weight. Thus, apart from thermal detailed-balance factors and convention-dependent constants, the rates scale as
\begin{equation}
\gamma_{mn}^{ab}(\omega)
\sim
J_{\rm local}^2\chi_{\mathrm{phys},ab}''(\rb_m-\rb_n,\omega).
\label{eq:rates_from_chi}
\end{equation}
This is the precise sense in which the same susceptibility that generates $\Hbus$ also sets the dissipative rates.

For a two-level qubit, it is useful to convert the frequency-resolved rate matrix $\gamma_{mn}^{ab}(\omega)$ into the physically named channels associated with the Bohr components $S^z$, $S^+$, and $S^-$. The transverse circular components describe spin-flip transitions, while the longitudinal component describes phase noise. We reserve uppercase $\Gamma$ for these named channel rates:
\begin{align}
\Gamma_{mn}^{\downarrow} &\equiv \gamma_{mn}^{+-}(\omega_0),
\label{eq:Gamma_down}
\\
\Gamma_{mn}^{\uparrow} &\equiv \gamma_{mn}^{-+}(-\omega_0),
\label{eq:Gamma_up}
\\
\Gamma_{mn}^{\phi} &\equiv \gamma_{mn}^{zz}(0).
\label{eq:Gamma_phi}
\end{align}
Numerical factors associated with the conversion between Cartesian and circular spin components are absorbed into the definitions of the channel rates in Eqs.~\eqref{eq:Gamma_down}--\eqref{eq:Gamma_phi}. Here $\Gamma^{\downarrow}$ is the downward transition rate: the qubit changes from its excited state to its ground state and emits energy $\omega_0$ into the bath. Conversely, $\Gamma^{\uparrow}$ is the upward transition rate: the qubit absorbs energy $\omega_0$ from the bath and is thermally excited from the ground state to the excited state. The longitudinal channel $\Gamma^\phi$ is the coefficient of the low-frequency longitudinal dissipator built from  $S^z$  operators and describes phase noise without changing the qubit population. With the spin normalization $S^z=\sigma^z/2$, the longitudinal dissipator with coefficient $\Gamma_{AA}^{\phi}$ contributes $\Gamma_{AA}^{\phi}/2$ to the decay rate of a single-qubit off-diagonal density-matrix element. The corresponding local coherence times therefore obey the standard relation between relaxation, longitudinal noise, and transverse coherence~\cite{BP2002book,Ithier2005}
\begin{equation}
\frac{1}{T_1}=\Gamma_{AA}^{\downarrow}+\Gamma_{AA}^{\uparrow},
\qquad
\frac{1}{T_2}=\frac{1}{2T_1}+\frac{1}{2}\Gamma_{AA}^{\phi}.
\label{eq:T1T2_defs}
\end{equation}
The first relation states that the population-relaxation rate is the sum of downward and upward transition rates. The second states that transverse coherence is lost both through population relaxation and through longitudinal phase noise. If instead one defines $\Gamma_{AA}^{\phi}$ as the physical pure-dephasing rate rather than the longitudinal Lindblad coefficient, the factor of $\frac{1}{2}$ is absorbed into that definition. Equations~\eqref{eq:Hbus_general} and \eqref{eq:rates_from_chi} therefore summarize the central message of the paper: the same physical susceptibility of the bath controls both the coherent mediated exchange and the decoherence generated by that bath.

\section{Mean-field Dirac-spinon benchmark}
\label{sec:mean_field}

The reduced open-system structure established above is completely general. Once the physical retarded susceptibility $\chiphys^R$ is specified, the bus Hamiltonian and the dissipative rates follow. To obtain explicit scaling laws, we now introduce a benchmark in which the physical spin susceptibility is approximated by the free Dirac-spinon bubble of the reference DSL theory. This approximation is not a solution of the full interacting gauge theory of the candidate $\mathrm{U(1)}$ DSL. It is an analytically transparent baseline against which the dressed susceptibility used in Sec.~\ref{sec:beyond_mean} can be compared~\cite{Hermele2004,Hermele2005,Ran2007,Hermele2008,Song2020PRX}.

\subsection{Benchmark bath Hamiltonian and spin density}

The four-flavor continuum theory in Eq.~\eqref{eq:L_dsl} may be organized as \(\alpha=(s,\lambda)\), where \(s=\uparrow,\downarrow\) labels physical spin and \(\lambda=1,2\) labels the two Dirac-node or valley flavors. At the spinon mean-field level, the emergent gauge field is fixed to its saddle-point background configuration, and fluctuations about that background are neglected. The bath is then approximated by neutral massless Dirac spinons with linear dispersion~\cite{Hermele2008,Hermele2004},
\begin{equation}
\begin{aligned}
H_{\mathrm{bath}}^{\mathrm{MF}}
&=
v_F\sum_{s=\uparrow,\downarrow}\sum_{\lambda=1}^{2}
\int \dd^2r\,
\\
&\quad\times
\psi_{s\lambda}^\dagger(\rb)
\left(-\ii\tau_x \partial_x-\ii\tau_y \partial_y\right)
\psi_{s\lambda}(\rb).
\end{aligned}
\label{eq:Hbath_MF}
\end{equation}
Here \(v_F\) is the Dirac spinon velocity, and \(\tau_{x,y}\) act in the two-component Dirac pseudospin sector.  Each field $\psi_{s\lambda}$ is a two-component Dirac spinor in this sector, while the labels $(s,\lambda)$ account for the four low-energy flavors inherited from Eq.~\eqref{eq:L_dsl}. This is the free-spinon benchmark associated with the four-flavor DSL field theory, not the full gauge-dressed response of the interacting spin liquid.

In this mean-field benchmark, the long-wavelength projection of the gauge-invariant bath spin operator is taken to be the fermion bilinear
\begin{equation}
 \mathcal{O}_{\mathrm{spin},\mathrm{MF}}^a(\rb)
=
\frac{1}{2}
\sum_{\lambda=1}^{2}
\sum_{s,s'=\uparrow,\downarrow}
\psi_{s\lambda}^\dagger(\rb)
\left(\sigma^a\right)_{ss'}
\psi_{s'\lambda}(\rb),
\label{eq:spin_density_MF}
\end{equation}
with an identity matrix understood in the Dirac pseudospin sector. Equation~\eqref{eq:spin_density_MF} is the simplest gauge-invariant spin-bilinear projection of the physical spin operator. More microscopic choices of the spin-operator projection change only the overall susceptibility amplitude used below~\cite{Hermele2005,Hermele2008,Song2019NatComm,Song2020PRX}.

The detailed one-loop derivation is given in Appendix~\ref{app:meanfield}. Here we quote only the resulting retarded benchmark susceptibility,
\begin{equation}
\chiMF^R(\qb,\omega)
=
-A_\chi\,
\frac{q^2}{\sqrt{v_F^2 q^2-(\omega+\ii0^{+})^2}},
\label{eq:chi0_ret}
\end{equation}
where \(q=|\qb|\), and \(A_\chi>0\) absorbs the flavor-counting convention, the normalization of the physical spin operator, and microscopic projection factors. The \(+\ii0^+\) prescription fixes the retarded branch of the square root. Equation~\eqref{eq:chi0_ret} is the spin-response analog of the massless two-dimensional Dirac polarization bubble~\cite{Wunsch2006,HwangDasSarma2007,Mahan2000}. It already contains the two kinematic features that matter most for the qubit problem, namely the nonanalytic static momentum dependence and the continuum threshold in the absorptive channel. This benchmark kernel thus provides the simplest analytically controllable realization of the general susceptibility-based framework developed in Sec.~\ref{sec:reduced_dynamics}. It is a benchmark input to the reduced-dynamics construction, not a claim that the full interacting $\mathrm{U(1)}$ spin liquid is exhausted by the free-bubble response.

\begin{figure}
\includegraphics[width=0.9\columnwidth]{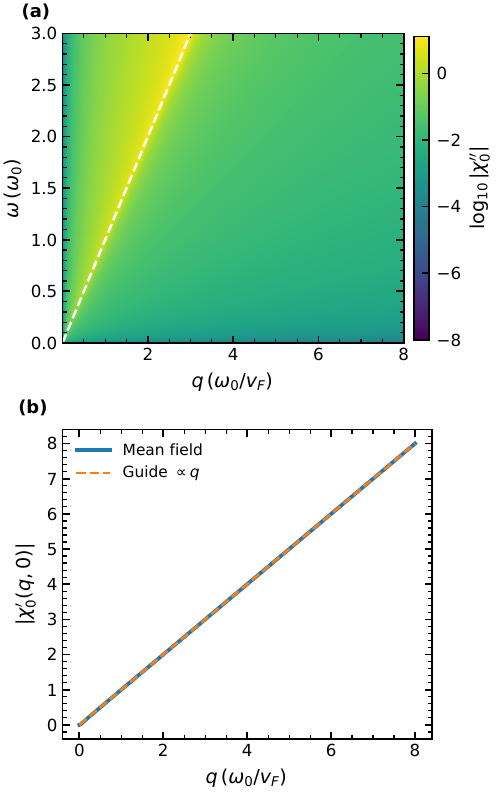}
\caption{Benchmark bath response in natural units. (a) Color map of \(\log_{10}|\chi_0''(q,\omega)|\), with momentum measured in units of \(\omega_0/v_F\) and frequency in units of \(\omega_0\). The dashed line marks the Dirac continuum threshold \(\omega=v_F q\). (b) Static benchmark response \(|\chi_{\mathrm{MF}}'(q,0)|\propto q\), showing the nonanalytic momentum dependence that underlies the algebraic real-space exchange coupling.}
\label{fig:bath_response}
\end{figure}


Figure~\ref{fig:bath_response} summarizes the numerical benchmark and highlights the two mean-field features that later control the reduced dynamics. The plot axes in the numerical figures are expressed in the natural units set by \(\omega_0\) and \(v_F\): momenta in units of \(\omega_0/v_F\), distances in units of \(v_F/\omega_0\), times in units of \(1/\omega_0\), and frequencies or rates in units of \(\omega_0\). Response amplitudes, susceptibilities, and derived rates are reported in the same scaled convention, rather than in separately fixed physical units. In Fig.~\ref{fig:bath_response}(a), the absorptive part \(\log_{10}|\chi_{\mathrm{MF}}''(\qb,\omega)|\) is concentrated above the Dirac continuum edge \(\omega=v_F q\), shown by the dashed line. The bath therefore does not behave as a featureless metallic reservoir at low frequency; instead, the dissipative phase space collapses as the qubit frequency is lowered, which is the microscopic origin of the pseudogap-protection argument used later for the local relaxation channel. Figure~\ref{fig:bath_response}(b) verifies the static benchmark law \(|\chi_{\mathrm{MF}}'(\qb,0)|\propto |\qb|\). Since the numerical plot is drawn only for radial momenta \((q\ge 0)\), this appears visually as a linear function of \(q\), but the essential point is the underlying nonanalytic Dirac form, because this static momentum dependence produces the algebraic real-space exchange discussed below. The mean-field kernel therefore serves as the analytically transparent baseline, while Sec.~\ref{sec:beyond_mean} introduces a single dressed-\(\mathrm{U}(1)\)-DSL-inspired deformation to test how strongly the reduced-dynamics conclusions depend on bath renormalization.

\subsection{Static algebraic exchange}

With the benchmark susceptibility in hand, we can now extract the coherent static interaction that serves as the reference bus law for the remainder of the paper. In the static limit,
\begin{equation}
\chiMF^R(\qb,0)
=
-A_\chi\frac{|\qb|}{v_F}.
\label{eq:chi0_static}
\end{equation}
Fourier transforming Eq.~\eqref{eq:chi0_static} to real space, as detailed in Appendix~\ref{app:meanfield}, gives the asymptotic form, consistent with the familiar \(R^{-3}\) behavior of undoped two-dimensional Dirac systems~\cite{Saremi2007,SherafatiSatpathy2011,Klinovaja2013},
\begin{equation}
\chiMF'(\Rb,0)
=
\frac{\mathcal C_\chi}{v_F R^3},
\qquad R\equiv \lvert\Rb\rvert \gg a,
\label{eq:chi0_realspace}
\end{equation}
where \(a\) is a microscopic cutoff and \(\mathcal C_\chi\) is a convention-dependent signed prefactor. Equations~\eqref{eq:chi0_static} and \eqref{eq:chi0_realspace} express the static benchmark reduction of the general bus formula of Sec.~\ref{sec:reduced_dynamics}. The nonanalytic Dirac law \(|\qb|\) in momentum space becomes an algebraic \(R^{-3}\) exchange kernel in real space. For the isotropic benchmark, we denote the corresponding radial response by \(\chiMF'(R,0)\). Hence the mediated exchange obeys
\begin{equation}
J_{\mathrm{eff}}(R)
=
J_{\rm local}^2\chiMF'(R,0)
\propto
\frac{J_{\rm local}^2}{v_F R^3},
\label{eq:Jeff_R3}
\end{equation}
up to an overall sign set by the microscopic exchange convention. Equation~\eqref{eq:Jeff_R3} is the algebraic bus law of the neutral Dirac benchmark and provides the reference scaling against which interaction-induced deviations may later be assessed.
\begin{figure*}
    \centering
    \includegraphics[width=\textwidth]{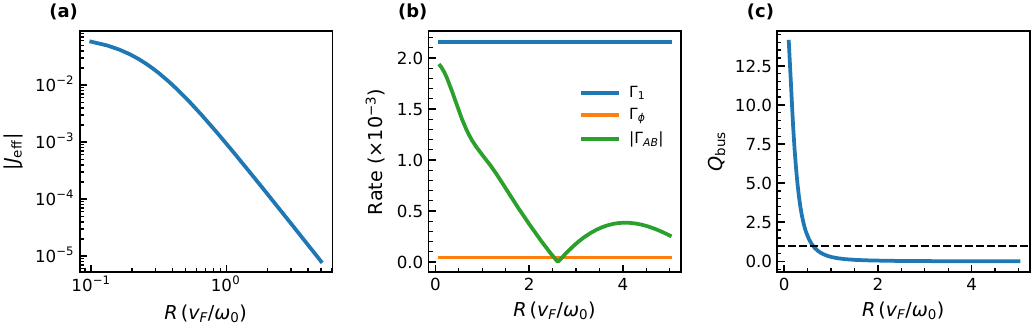}
  \caption{
Benchmark coherent and dissipative scales. 
(a) Static bath-mediated exchange \(J_{\mathrm{eff}}(R)\), with \(R\) measured in units of \(v_F/\omega_0\).  (b) Local relaxation \(\Gamma_1\), local pure-dephasing coefficient \(\Gamma_\phi\), and nonlocal cross-damping magnitude \(|\Gamma_{AB}|\) in the same dimensionless benchmark units. (c) Bus metric \(Q_{\mathrm{bus}}(R)\); the dashed line marks the nominal viability threshold \(Q_{\mathrm{bus}}=1\).
}
    \label{fig:exchange_rates}
\end{figure*}


\subsection{Pseudogap-suppressed relaxation}

The absorptive part of Eq.~\eqref{eq:chi0_ret} is nonzero only inside the particle-hole continuum,
\begin{equation}
|\chiMF''(\qb,\omega)|
=
A_\chi\,
\frac{q^2}{\sqrt{\omega^2-v_F^2 q^2}}
\Theta(\omega-v_F q),
\label{eq:chi0_imag}
\end{equation}
for \(\omega>0\), where $q=|\qb|$.  Here the absolute value denotes the positive absorptive spectral weight; the overall sign of $\chiMF''$ depends on the retarded-response convention. Equation~\eqref{eq:chi0_imag} makes the kinematic origin of bath-induced dissipation explicit. For a fixed momentum $q$, the bath can absorb qubit energy only when the threshold condition $\omega>v_F q$ is satisfied. Because a local qubit couples at a point, it samples all bath momenta. At low temperature and for positive transition frequency, the ordered local transverse spectrum is obtained, up to convention-dependent thermal and normalization factors, by integrating the positive-frequency absorptive spectral weight over $\qb$:
\begin{equation}
\mathcal{S}_{\perp\perp}^{>}(0,\omega_0)
\sim
\int \frac{\dd^2 q}{(2\pi)^2}\,|\chiMF''(\qb,\omega_0)|
\propto
\frac{\omega_0^3}{v_F^4}.
\label{eq:S_local_scaling}
\end{equation}
The corresponding downward relaxation scale therefore obeys
\begin{equation}
\Gamma_1 \equiv \Gamma_{AA}^{\downarrow}
\propto
J_{\rm local}^2\frac{\omega_0^3}{v_F^4}.
\label{eq:Gamma1_scaling}
\end{equation}
Equations~\eqref{eq:S_local_scaling} and \eqref{eq:Gamma1_scaling} are the dissipative counterpart of the static exchange law above. The same benchmark susceptibility that produces an algebraic coherent bus also yields a strongly suppressed positive-frequency relaxation channel, reflecting the pseudogapped low-energy phase space of massless Dirac excitations~\cite{Wunsch2006,HwangDasSarma2007,Clerk2010RMP}. In this sense, the neutral Dirac bath offers pseudogap protection: it is not noiseless, but low-frequency downward-relaxation phase space collapses rapidly as $\omega_0\to 0$.
\subsection{Operational bus quality}

Combining Eqs.~\eqref{eq:Jeff_R3} and \eqref{eq:Gamma1_scaling} gives a simple relaxation-limited estimate,
\begin{equation}
\Qbus^{(0)}(R)
\equiv
\frac{|J_{\mathrm{eff}}(R)|}{\Gamma_1}
\sim
\left(\frac{v_F}{\omega_0 R}\right)^3.
\label{eq:Qbus_scaling}
\end{equation}
This estimate isolates the analytic Dirac scaling obtained from the static exchange and the local downward-relaxation channel. It also identifies the dynamic length scale
\begin{equation}
\ell_\omega \sim \frac{v_F}{\omega_0},
\label{eq:ellomega}
\end{equation}
with a natural operating regime  \(R\ll \ell_\omega\) in which coherent nonlocal exchange outpaces local relaxation. Thus, even before including the additional dissipative channels used in the numerical analysis, the benchmark identifies the principal spatial and spectral scales governing the usefulness of the bath as an entanglement bus.

\begin{figure}[t]
    \includegraphics[width=0.9\columnwidth]{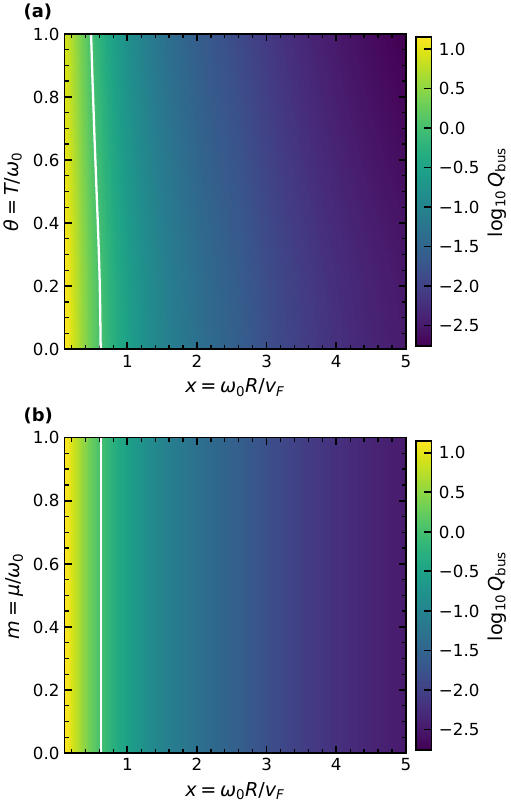}
    \caption{Baseline operating-window map for the benchmark bath. 
(a) \(\log_{10}Q_{\rm bus}(x,\theta)\) as a function of the dimensionless separation \(x=\omega_0R/v_F\) and the dimensionless temperature \(\theta=T/\omega_0\), evaluated at charge neutrality. 
(b) \(\log_{10}Q_{\rm bus}(x,m)\) as a function of \(x\) and the dimensionless chemical potential \(m=\mu/\omega_0\), evaluated at the reference low-temperature point. 
The white contour marks the nominal boundary \(Q_{\rm bus}=1\), separating the exchange-dominated region from the dissipation-dominated region within the benchmark model.}
    \label{fig:phase_map}
\end{figure}

The numerical figures use a slightly more conservative operational metric that also includes longitudinal dephasing and nonlocal cross-damping. These benchmark quantities are the direct input to Figs.~\ref{fig:exchange_rates} and \ref{fig:phase_map}. Figure~\ref{fig:exchange_rates} is the point at which the bath susceptibility is reduced to the concrete quantities that enter the effective master equation. Figure~\ref{fig:exchange_rates}(a) reports the smoothed static benchmark exchange \(J_{\mathrm{eff}}(R)\) extracted from the real-space kernel. Figure~\ref{fig:exchange_rates}(b) separates the dissipative inputs into a local single-qubit downward relaxation scale \(\Gamma_1\equiv \Gamma_{AA}^{\downarrow}\), a local longitudinal-dephasing coefficient \(\Gamma_\phi\equiv \Gamma_{AA}^{\phi}\), and the nonlocal cross-damping magnitude \(|\Gamma_{AB}|\). In the present benchmark implementation, the first two are treated as local bath properties and therefore appear as \(R\)-independent baselines, whereas \(\Gamma_{AB}\) is extracted from the nonlocal imaginary part of the real-space susceptibility and retains explicit spatial dependence.

Figure~\ref{fig:exchange_rates}(c) assembles these ingredients into the operational bus metric
\begin{equation}
\Qbus(R)
=
\frac{|J_{\mathrm{eff}}(R)|}
{\Gamma_1+\Gamma_\phi+|\Gamma_{AB}(R)|},
\label{eq:Qbus_numeric_definition}
\end{equation}
so that the reference line \(\Qbus=1\) gives a simple boundary between exchange-dominated and dissipation-dominated behavior. Figure~\ref{fig:phase_map} then turns the same operational metric into an operating-window plot in the dimensionless variables \(x\equiv \omega_0R/v_F\), \(\theta\equiv T/\omega_0\), and \(m\equiv \mu/\omega_0\). Here \(x\) measures the qubit separation in units of the dynamical length \(v_F/\omega_0\), while \(\theta\) and \(m\) measure the thermal and chemical-potential scales relative to the qubit splitting. In the present benchmark, the contours are controlled primarily by \(x\), reflecting the dominant role of the exchange-to-dissipation balance as the qubits are separated. The phase map therefore serves as a baseline operating-window plot for the bus mechanism, rather than a material-specific optimization map.
\section{Beyond mean field: dressed physical susceptibility}
\label{sec:beyond_mean}
The mean-field benchmark of Sec.~\ref{sec:mean_field} was introduced only as the simplest analytically controllable realization of the bath kernel. A full $\mathrm{U(1)}$ DSL, however, is an interacting gauge theory rather than a free spinon gas. The mean-field result therefore captures only the baseline Dirac kinematics: in the full problem, both the propagating spinons and the operator that probes physical spin are dressed by gauge fluctuations and other interaction effects~\cite{Hermele2005,Hermele2008,Song2019NatComm,Song2020PRX}. Accordingly, we treat the bath kernel entering the reduced qubit dynamics as a dressed gauge-invariant susceptibility, not as the bare spinon bubble. Schematically, the dressed-bubble contribution to the physical susceptibility may be written as
\begin{equation}
\begin{split}
\chi_{\mathrm{phys}}^{R,ab}(\qb,\omega)
={}&
-\ii \int \frac{\dd\nu\,\dd^2\kb}{(2\pi)^3}
\Tr\Big[
\mathcal{V}^a(\kb+\qb,\kb;\nu+\omega,\nu)
\\
&\qquad \times G(\kb+\qb,\nu+\omega)
\\
&\qquad \times \mathcal{V}^b(\kb,\kb+\qb;\nu,\nu+\omega)
\\
&\qquad \times G(\kb,\nu)
\Big]
+\chi_{\mathrm{other}}^{R,ab}(\qb,\omega).
\end{split}
\label{eq:chi_phys_formal}
\end{equation}
In Eq.~\eqref{eq:chi_phys_formal}, \(\qb\) is the external momentum carried by the spin response, while \(\kb\) is the internal loop momentum integrated over in the dressed bubble. The integral is over one frequency and two spatial momenta, appropriate to the \(2+1\)-dimensional DSL. Here \(G\) is the dressed spinon propagator, \(\mathcal{V}^a\) is the dressed physical-spin vertex, and \(\chi_{\mathrm{other}}^{R,ab}\) denotes additional gauge-invariant critical contributions not exhausted by the dressed bubble channel.

The key structural point is that the reduced qubit-sector derivation does not change. All of the open-system machinery in Sec.~\ref{sec:reduced_dynamics} is written in terms of the physical susceptibility and therefore applies equally to the mean-field benchmark and to a dressed interacting kernel. In practice, the most convenient route is to preserve the same qubit-sector mapping and replace the benchmark kernel $\chi_0^R$ by a dressed physical susceptibility $\chiphys^R$.

\begin{figure*}[t]
    \centering
    \includegraphics[width=\textwidth]{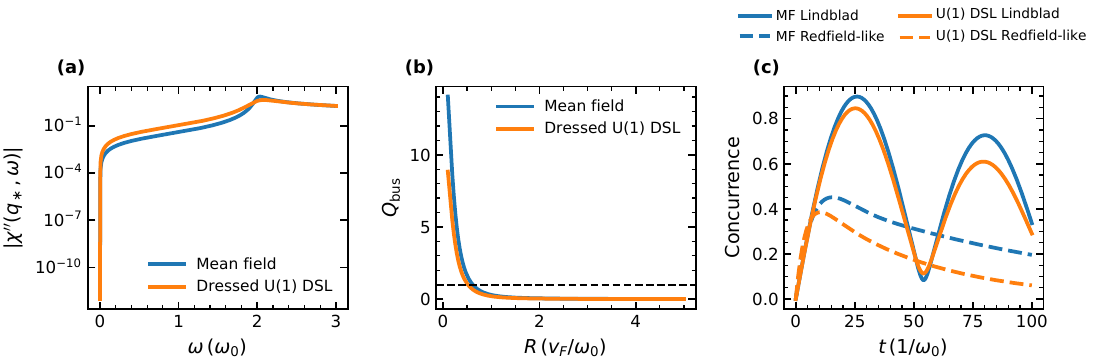}
\caption{Robustness of the spin-bus mechanism beyond the mean-field benchmark. 
(a) Kernel-level comparison at fixed momentum \(q_\ast\), showing the absorptive spectral cut \(|\chi''(q_\ast,\omega)|\) with \(\omega\) measured in units of \(\omega_0\). 
(b) Bus-level comparison of the corresponding \(Q_{\rm bus}(R)\) profiles, with \(R\) measured in units of \(v_F/\omega_0\). The dressed bath preserves a finite exchange-dominated regime but shifts the viability boundary toward smaller \(R\). 
(c) Representative time-domain comparison of the concurrence in the selected operating sector, with \(t\) measured in units of \(1/\omega_0\), showing mean-field and dressed-\(\mathrm{U}(1)\)-DSL-inspired Lindblad and Redfield-like dynamics.
}
    \label{fig:robustness}
\end{figure*}

At the level of static scaling, motivated by the anomalous power laws of algebraic spin-liquid/QED$_3$ response functions~\cite{Hermele2005,Hermele2008,ChesterPufu2016}, we take the long-wavelength ansatz
\begin{equation}
\chiphys^{R}(\qb,0)
\sim
Z_\chi\,q^{1-\eta_s},
\label{eq:chi_scaling_eta}
\end{equation}
which implies, in real space,
\begin{equation}
\chiphys^{R}(\Rb,0)
\sim
\frac{1}{R^{3-\eta_s}}.
\label{eq:chi_realspace_eta}
\end{equation}
Equation~\eqref{eq:chi_realspace_eta} is the direct beyond-mean-field analog of the $R^{-3}$ benchmark law of Sec.~IV: the coherent bus remains algebraic, but its exponent is renormalized by the dressed bath response.

For the numerical simulations, we use a concrete dressed-$\mathrm{U(1)}$-DSL-inspired retarded kernel obtained by deforming the benchmark Dirac form,
\begin{equation}
\chiphys^R(\qb,\omega)
=
-\,Z_\chi A_\chi\,
\frac{q^{\,2-\eta_s}}
{\sqrt{v_F^2 q^2-\bigl(\omega+\ii\gamma_{\mathrm{eff}}\bigr)^2}},
\label{eq:chi_u1dsl_numeric}
\end{equation}
with
\begin{equation}
\gamma_{\mathrm{eff}}
=
\eta_{\mathrm{reg}}
+
\gamma_{\mathrm{broad}}
+
c_T\,T
+
c_\mu\,|\mu|.
\label{eq:gamma_eff_numeric}
\end{equation}
Equation~\eqref{eq:chi_u1dsl_numeric} is intended as a broadened numerical ansatz for robustness studies. When $\gamma_{\rm eff}$ is nonzero, it smooths the strict low-momentum asymptotic power law below $q\sim \gamma_{\rm eff}/v_F$; the algebraic form in Eqs.~\eqref{eq:chi_scaling_eta} and \eqref{eq:chi_realspace_eta} then represents the underlying scaling limit recovered when the broadening is negligible in the static sector. Here $Z_\chi$ renormalizes the overall spectral weight, $\eta_s$ allows the low-momentum response to deviate from the strict mean-field power law, and $\gamma_{\mathrm{eff}}$ broadens the continuum edge in a way that can depend on temperature and doping. The purpose of Eq.~\eqref{eq:chi_u1dsl_numeric} is not to provide a complete microscopic theory of the interacting spin liquid, but to give a transparent numerical ansatz that mimics, at a phenomenological level, the type of reshaping expected once gauge-field and interaction effects are no longer ignored.

This single dressed-$\mathrm{U(1)}$-DSL-inspired kernel is used throughout the robustness analysis. We do not include a separate amplitude-only comparison, \(\chiMF^R\to Z_\chi\chiMF^R\), since such a rescaling would leave both the low-momentum power law and the continuum broadening unchanged. The more relevant test is whether a structured beyond-mean-field deformation, combining modified low-momentum scaling with enhanced broadening, preserves the operating-window picture. Equations~\eqref{eq:Hbus_general}--\eqref{eq:rates_from_chi} therefore remain unchanged, with only the bath input replaced by the dressed susceptibility. A compact summary of the numerical parameterization is given in Appendix~\ref{app:dressed}.

Figure~\ref{fig:robustness} shows the numerical consequence of this replacement. Panel~(a) compares the benchmark spectral cut with the dressed-$\mathrm{U(1)}$-DSL-inspired kernel at fixed momentum, showing that dressing reshapes the absorptive response without removing the underlying Dirac structure. Panel~(b) compares the corresponding bus metrics: a finite exchange-dominated regime remains, but the \(Q_{\mathrm{bus}}=1\) boundary shifts toward smaller \(R\). Panel~(c) gives a representative time-domain comparison in the exchange-dominated sector, where the dressed bath still permits coherent entanglement generation but with a reduced coherent margin and stronger damping. The figure therefore functions as a robustness test of the same susceptibility-based mechanism, not as a second independent benchmark.
\section{Reduced-dynamics simulations}
\label{sec:sim}

With the bath kernel specified, the remaining task is purely at the qubit level: one solves the reduced master equation using the coherent and dissipative coefficients extracted from the chosen susceptibility~\cite{BP2002book,Johansson2012,Johansson2013}. The simulation stage is thus the final step of the modular framework developed in Secs.~II--V. The bath enters only through the effective kernels, while the qubit-sector dynamics is solved independently of the microscopic route used to obtain them.
Once the effective bath kernel has been specified, the reduced dynamics is solved from
\begin{equation}
\dot\rho
=
-\ii[\Heff,\rho]
+\Dcal[\rho],
\label{eq:effective_master}
\end{equation}
where the effective Hamiltonian is
\begin{equation}
\Heff
=
\frac{\omega_0}{2}\sigma_A^z
+
\frac{\omega_0}{2}\sigma_B^z
+
\HLS^{\mathrm{loc}}
+
\Hbus.
\label{eq:Heff_sim}
\end{equation}
Here $\HLS^{\mathrm{loc}}$ collects the local coherent renormalizations inherited from the bath, while $\Hbus$ is the nonlocal interaction kernel derived in Sec.~III. In the isotropic case,
\begin{equation}
\Hbus
=
J_{\mathrm{eff}}(R)\,\Sb_A\cdot\Sb_B,
\label{eq:Hbus_sim}
\end{equation}
while the dissipator may be expressed, in the Markovian weak-coupling limit, as
\begin{align}
\Dcal[\rho]
=&
\sum_{m,n}
\Gamma_{mn}^{\downarrow}
\left(S_n^-\rho S_m^+ - \frac{1}{2}\{S_m^+S_n^-,\rho\}\right)
\nonumber \\
&+
\sum_{m,n}
\Gamma_{mn}^{\phi}
\left(S_n^z\rho S_m^z - \frac{1}{2}\{S_m^zS_n^z,\rho\}\right).
\label{eq:dissipator_sim}
\end{align}
In Eq.~\eqref{eq:dissipator_sim}, $\Gamma_{mn}^{\phi}$ is the coefficient of the longitudinal $S^z$ dissipator. With $S^z=\sigma^z/2$, the corresponding local contribution to the physical single-qubit pure-dephasing rate is $\Gamma_{AA}^{\phi}/2$, consistent with Eq.~\eqref{eq:T1T2_defs}.

\begin{figure*}
    \centering
    \includegraphics[width=0.98\textwidth]{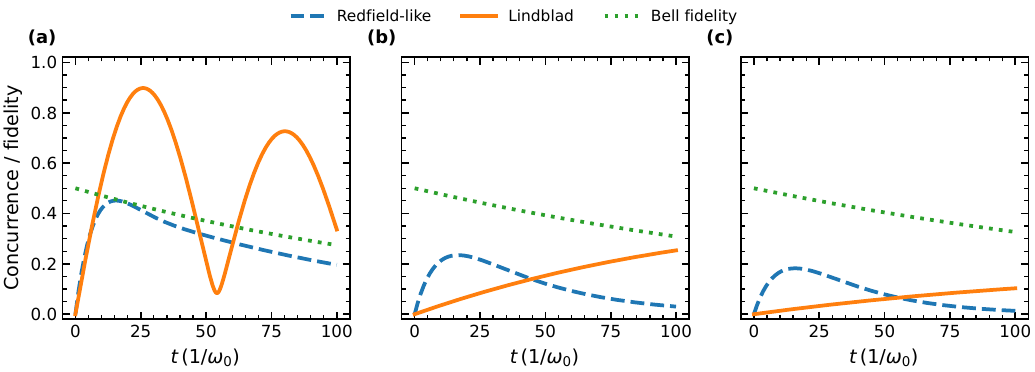}
    \caption{
Representative reduced dynamics at the three operating points selected from Fig.~\ref{fig:phase_map}. 
(a) Deep-viable sector \((Q_{\mathrm{bus}}\gg 1)\). 
(b) Boundary sector \((Q_{\mathrm{bus}}\approx 1)\). 
(c) Outside sector \((Q_{\mathrm{bus}}<1)\). 
Time is measured in units of \(1/\omega_0\). Solid curves show Lindblad dynamics, dashed curves show Redfield-like dynamics, and the dotted curve shows the Bell-state fidelity.
}
    \label{fig:dynamics}
\end{figure*}


The simulation inputs are therefore $J_{\mathrm{eff}}(R)$, $\Gamma_{mn}^{\downarrow}$, and $\Gamma_{mn}^{\phi}$, all extracted from the chosen bath kernel. If thermal excitation or additional collective dissipative channels are retained, the corresponding $\Gamma_{mn}^{\uparrow}$ and off-diagonal rate tensors are included straightforwardly through the same formalism. A compact summary of the numerical pipeline and observables is collected in Appendix~\ref{app:numerics}.

For the numerical results shown below, we use dimensionless benchmark units with $\omega_0=1$ and $v_F=1$. Frequencies and rates are therefore measured in units of the qubit splitting, while distances are measured in units of $v_F/\omega_0$. The local coupling is set to $J_{\rm local}=0.08$, placing the simulations in the weak-coupling regime where the induced exchange and dissipative rates scale as $J_{\rm local}^2$. The reference temperature is specified by the dimensionless ratio $\theta\equiv  T/\omega_0=0.05$, corresponding to a low-temperature benchmark with $\beta\omega_0=20$, and the chemical potential is set to $m\equiv\mu/\omega_0=0$ to represent the neutral Dirac point. These values are not intended as material-specific parameters; they define a representative baseline for illustrating the susceptibility-to-dynamics pipeline, while the operating-window plots vary the dimensionless control variables $x=\omega_0R/v_F$, $\theta$, and $m$. 
The bath kernel is sampled on finite $(q,\omega)$ grids and regularized by a small numerical broadening so that the dynamic response shown in Fig.~\ref{fig:bath_response}(a) remains smooth enough to define stable local spectra and real-space exchange estimates. The operating points used in the reduced-dynamics figures are selected directly from the bus metric: a deep-viable point with $\Qbus\gg 1$, a boundary point with $\Qbus\approx 1$, and an outside point with $\Qbus<1$. Thus the time-domain figures are not independent simulations at arbitrary distances, but dynamical probes of the three operating sectors identified in Fig.~\ref{fig:phase_map}.

A useful diagnostic timescale is the coherent exchange scale
\begin{equation}
t_{\rm ex}\sim \frac{1}{|J_{\mathrm{eff}}(R)|},
\label{eq:tex}
\end{equation}
which estimates how rapidly the mediated interaction can generate nonlocal two-qubit dynamics, up to protocol-dependent numerical factors. The reduced dynamics is then assessed using the single-qubit coherence times from Eq.~\eqref{eq:T1T2_defs}, the concurrence \(C[\rho(t)]\), Bell-state fidelity, purity \(\Tr\rho^2(t)\), and the polarization histories \(\langle\sigma_A^z(t)\rangle\) and \(\langle\sigma_B^z(t)\rangle\). The use of concurrence follows the standard two-qubit entanglement measure of Wootters~\cite{Wootters1998}, while fidelity and purity are standard state-diagnostic quantities in quantum information~\cite{NielsenChuang2010}. Numerically, we use two complementary reduced-dynamics implementations. The first is a direct Lindblad evolution with fixed effective rates extracted from the susceptibility-derived kernels. The second is a Redfield-like evolution in which the frequency dependence and collective local--nonlocal structure of the effective bath spectral tensor are retained more explicitly~\cite{BP2002book,Johansson2012,Johansson2013}. The useful feature of the present formulation is that the simulation pipeline is decoupled from the microscopic route used to obtain the bath kernel.  The same reduced-dynamics workflow can be run with the mean-field kernel $\chiMF^R$ or with a dressed physical susceptibility $\chiphys^R$.

The main reduced-dynamics comparison is shown in Fig.~\ref{fig:dynamics}, where the Lindblad benchmark and the Redfield-like evolution are evaluated at the three operating points selected from Fig.~\ref{fig:phase_map}, with the Lindblad and Redfield dynamics implemented numerically using QuTiP~\cite{Johansson2012,Johansson2013}. The figure provides a dynamical readout of the operating-window analysis, not an independent set of numerical examples. In the deep-viable regime, where the benchmark metric satisfies $\Qbus\gg 1$, the Lindblad curves display the clearest exchange-driven buildup of two-qubit correlations, while the Redfield-like evolution shows the stronger damping that appears when the susceptibility-derived bath spectrum is retained beyond a fixed-rate Lindblad approximation. Near the boundary, coherent exchange and dissipation are comparable, so the time traces become more weakly oscillatory and the entanglement becomes transient. Outside the window, where $\Qbus<1$, both solvers show that correlations are weak or short-lived on the simulated timescale. The same dataset is unpacked further in Supplemental Material, where concurrence, Bell fidelity, purity, and polarization histories are shown separately so that coherent oscillatory exchange can be distinguished from bath-driven transients.

\section{Discussion and conclusion}
\label{sec:diss}

We have formulated the entanglement-bus problem for two localized qubits coupled through a candidate $\mathrm{U(1)}$ Dirac-spin-liquid-like bath in a way that cleanly separates the many-body bath problem from the reduced qubit dynamics. The qubit-sector derivation depends only on the physical retarded spin susceptibility of the bath. Once this quantity is specified, the same framework yields both the coherent bus Hamiltonian and the decoherence rates. This is the central conceptual result of the paper.

A further motivation comes from geometry and scaling. Direct wave-function overlap typically falls exponentially with qubit separation, whereas conventional quantum interconnects rely on engineered electromagnetic, photonic, phononic, or resonator modes to mediate longer-range coupling~\cite{Blais2004,Majer2007,Kurizki2015,Gustafsson2014,Satzinger2018}. The DSL-based route considered here occupies a different regime, using a two-dimensional correlated spin medium whose coupling and noise are both determined by the many-body spin susceptibility. Such a planar bath is naturally compatible with van der Waals and other atomically thin heterostructure layouts, where qubits or quantum emitters can be hosted in nearby layers~\cite{Onizhuk2021,Srivastava2015,Tran2016,Dastidar2022}. This motivates the central question addressed in this work, namely whether susceptibility-mediated exchange can dominate over the decoherence generated by the same bath.

The neutral Dirac-spinon mean-field theory provides the simplest controlled setting in which to answer this question. It furnishes explicit, analytically transparent scaling laws for the coherent and dissipative kernels. In particular, it gives the algebraic exchange $J_{\rm eff}(R)\propto J_{\rm local}^2/(v_F R^3)$ together with pseudogap-suppressed relaxation $\Gamma_1\propto J_{\rm local}^2\omega_0^3/v_F^4$. These relations define a baseline operating window and identify the dynamic length scale $\ell_\omega\sim v_F/\omega_0$ as the natural radius of the coherent bus.

Beyond mean field, gauge-field dressing and other interaction effects do not invalidate the reduced qubit-sector theory; they enter only through the replacement of the benchmark susceptibility by the physical gauge-invariant susceptibility of the interacting DSL. This makes the framework readily extensible to microscopic calculations or phenomenological kernels appropriate to candidate correlated materials.

Taken together, Figs.~\ref{fig:bath_response}--\ref{fig:dynamics} support a bounded but positive verdict on the $\mathrm{U(1)}$-DSL bus. The benchmark susceptibility furnishes a well-defined algebraic exchange law and a pseudogap-suppressed local relaxation channel; the bus metric and phase map then show that these ingredients generate a finite exchange-dominated operating window rather than an unbounded low-loss regime; and the reduced-dynamics simulations confirm that this window corresponds to genuinely different two-qubit behavior in the deep-viable, boundary, and outside regimes. The dressed-kernel comparison further shows that beyond-mean-field renormalization changes the quantitative size of the window without eliminating the mechanism itself.

The appendices are organized to mirror the logic of the main text. Appendix~\ref{app:redfield} records the frequency-resolved Born--Markov--secular derivation underlying Sec.~\ref{sec:reduced_dynamics},  Appendix~\ref{app:meanfield} supplies the one-loop benchmark susceptibility, and Appendices~\ref{app:dressed} and \ref{app:numerics} collect the dressed-kernel parameterization and simulation details. This modular structure is intended to make the framework useful both as a conceptual description of a spin-liquid entanglement bus and as a practical starting point for reduced-dynamics calculations in more microscopic models.


\appendix

\section{Born-Markov derivation of the reduced master equation}
\label{app:redfield}

Starting from Eq.~\eqref{eq:born_markov_start}, we write the interaction-picture Hamiltonian as in Eq.~\eqref{eq:Hint_t} and insert it into the double commutator. This generates terms of the form
\begin{equation}
\sum_{m,n}\sum_{a,b}
S_m^a(t)S_n^b(t-\tau)
\,C_{mn}^{ab}(\tau),
\label{eq:double_comm_terms}
\end{equation}
with $C_{mn}^{ab}(\tau)$ defined in Eq.~\eqref{eq:bath_corr_matrix}. Decomposing the system operators into Bohr-frequency components as in Eq.~\eqref{eq:bohr_decomp}, one obtains
\begin{equation}
S_m^a(t)=\sum_\omega \ee^{-\ii\omega t}S_m^a(\omega).
\label{eq:Smw_appendix}
\end{equation}
The one-sided Fourier transform of the bath correlations is
\begin{equation}
\begin{aligned}
\mathcal{K}_{mn}^{ab}(\omega)
&=
J_{\rm local}^2 \int_0^\infty \dd\tau\;
\ee^{\ii\omega\tau} C_{mn}^{ab}(\tau)
\\
&=
\frac{1}{2}\gamma_{mn}^{ab}(\omega)
+\ii\Lambda_{mn}^{ab}(\omega).
\end{aligned}
\label{eq:Gamma_decomp}
\end{equation}
Here $\gamma_{mn}^{ab}(\omega)$ is the dissipative Hermitian rate matrix, while $\Lambda_{mn}^{ab}(\omega)$ parameterizes the principal-value part that enters the coherent Lamb-shift and exchange terms. After the usual Born--Markov reduction followed by the secular organization in the Bohr frequencies~\cite{Redfield1957,Davies1974,BP2002book}, the master equation becomes
\begin{equation}
\begin{aligned}
\dot\rho
={}&
-\ii[H_q+\HLS,\rho]
\\
&+
\sum_\omega\sum_{m,n}\sum_{a,b}
\gamma_{mn}^{ab}(\omega)
\Big[
S_n^b(\omega)\rho S_m^{a\dagger}(\omega)
\\
&\hspace{2.2cm}
-\frac{1}{2}
\bigl\{
S_m^{a\dagger}(\omega)S_n^b(\omega),\rho
\bigr\}
\Big].
\end{aligned}
\label{eq:master_appendix}
\end{equation}
with
\begin{equation}
\HLS
=
\sum_\omega\sum_{m,n}\sum_{a,b}
\Lambda_{mn}^{ab}(\omega)
S_m^{a\dagger}(\omega)S_n^b(\omega).
\label{eq:HLS_appendix}
\end{equation}
The nonlocal bus Hamiltonian is the $m\neq n$ piece of Eq.~\eqref{eq:HLS_appendix}; its tensor structure reduces to Eq.~\eqref{eq:Hbus_general} in the isotropic case. Likewise, the dissipative rates are the corresponding real parts at $\omega=\pm\omega_0$ and $\omega=0$, giving Eqs.~\eqref{eq:Gamma_down}--\eqref{eq:Gamma_phi}.

\section{Mean-field Dirac-spinon benchmark susceptibility}
\label{app:meanfield}

In this appendix we derive the mean-field Dirac-spinon susceptibility quoted in Sec.~\ref{sec:mean_field}. The calculation is the free-bubble benchmark associated with the four-flavor DSL field theory of Eq.~\eqref{eq:L_dsl}. Physical spin gives a twofold spin structure, and the two Dirac nodes give a valley degeneracy \(N_v=2\). The result is the spin-response analog of the standard massless two-dimensional Dirac polarization bubble~\cite{Wunsch2006,HwangDasSarma2007,Mahan2000}.

We begin from the imaginary-time spin correlator
\begin{equation}
\mathcal{X}_{ab}(\rb,\tau)
=
\ev{T_\tau s^a(\rb,\tau)s^b(0,0)}_{\mathrm{MF}}.
\label{eq:Xab_tau}
\end{equation}
Its Fourier transform is
\begin{equation}
\mathcal{X}_{ab}(\qb,\ii\Omega_n)
=
\int_0^\beta \dd\tau\int \dd^2r\;\ee^{\ii(\Omega_n\tau-\qb\cdot\rb)}\mathcal{X}_{ab}(\rb,\tau).
\label{eq:Xab_qw}
\end{equation}
The retarded benchmark kernel quoted in the main text is obtained from this Matsubara correlator by analytic continuation after the one-loop evaluation described below.

Using the bilinear spin density of Eq.~\eqref{eq:spin_density_MF} and Wick's theorem, the connected correlator reduces to the one-loop bubble
\begin{equation}
\begin{aligned}
\mathcal{X}_{ab}(\qb,\ii\Omega_n)
&=
-\frac{1}{4\beta}
\sum_{\nu_n}
\int\frac{\dd^2k}{(2\pi)^2}
\\
&\quad\times
\Tr\!\Big[
\mathcal{V}_0^a
G_0(\kb+\qb,\ii\nu_n+\ii\Omega_n)
\\
&\qquad\quad\times
\mathcal{V}_0^b
G_0(\kb,\ii\nu_n)
\Big].
\end{aligned}
\label{eq:bubble_start}
\end{equation}
Here $\mathcal V_0^a=\sigma^a\otimes I_\lambda\otimes I_\tau$ is the bare mean-field spin vertex. The overall prefactor $1/4$ in Eq.~\eqref{eq:bubble_start}already accounts for the two factors of $1/2$ coming from the bilinear spin-density operator in Eq.~\eqref{eq:spin_density_MF} . The trace in Eq.~\eqref{eq:bubble_start}  is over physical spin, valley, and Dirac pseudospin. The free propagator is diagonal in physical spin and valley space,

\begin{equation}
\begin{aligned}
G_0(\kb,\ii\nu_n)
&=
I_\sigma\otimes I_\lambda\otimes g_0(\kb,\ii\nu_n),
\\
g_0(\kb,\ii\nu_n)
&=
\frac{\ii\nu_n+v_F\bm{\tau}\cdot\kb}
{(\ii\nu_n)^2-v_F^2k^2}.
\end{aligned}
\label{eq:G0_g0}
\end{equation}
The trace over physical spin gives \(\Tr_\sigma[\sigma^a\sigma^b]=2\delta_{ab}\), while the valley trace gives \(N_v=2\). The susceptibility is therefore isotropic,
\begin{equation}
\mathcal{X}_{ab}(\qb,\ii\Omega_n)=\delta_{ab}\,\mathcal{X}(\qb,\ii\Omega_n).
\label{eq:X_isotropic}
\end{equation}
The remaining Dirac pseudospin trace yields
\begin{equation}
\begin{aligned}
\mathcal{X}(\qb,\ii\Omega_n)
&=
-N_v
\int\frac{\dd\nu}{2\pi}
\int\frac{\dd^2k}{(2\pi)^2}
\\
&\frac{
-\nu(\nu+\Omega_n)+v_F^2\kb\cdot(\kb+\qb)
}{
\left[(\nu+\Omega_n)^2+v_F^2|\kb+\qb|^2\right]
\left[\nu^2+v_F^2k^2\right]
},
\end{aligned}
\label{eq:X_integral}
\end{equation}
with \(N_v=2\) for the four-flavor spin-$1/2$ DSL convention. The precise overall factor depends on the normalization of the physical spin operator; in the main text it is absorbed into \(A_\chi\).

Applying a Feynman-parameter decomposition,
\begin{equation}
\frac{1}{AB}=\int_0^1 \dd x\;\frac{1}{[xA+(1-x)B]^2},
\label{eq:feynman_param}
\end{equation}
shifting the loop momentum, and performing the frequency and momentum integrations gives
\begin{equation}
\mathcal{X}(\qb,\ii\Omega_n)
=
A_\chi\,
\frac{q^2}{\sqrt{v_F^2 q^2+\Omega_n^2}},
\label{eq:X_qw_final}
\end{equation}
where \(A_\chi>0\) absorbs the valley/flavor factor, the spin-operator normalization, and microscopic projection factors. Analytic continuation \(\ii\Omega_n\to \omega+\ii0^{+}\) gives the retarded benchmark susceptibility,
\begin{equation}
\chiMF^R(\qb,\omega)
=
-A_\chi\,
\frac{q^2}{\sqrt{v_F^2 q^2-(\omega+\ii0^{+})^2}}.
\label{eq:chi0_ret_appendix}
\end{equation}

The static limit is
\begin{equation}
\chiMF^R(\qb,0)
=
-A_\chi\frac{|\qb|}{v_F}.
\label{eq:chi0_static_appendix}
\end{equation}
Using rotational symmetry, its real-space transform may be written formally as
\begin{equation}
\chiMF'(\Rb,0)
=
-\frac{A_\chi}{2\pi v_F}
\int_0^\infty \dd q\; q^2 J_0(qR),
\label{eq:chiR_integral_real}
\end{equation}
where \(R=|\Rb|\), and \(J_0\) comes from the angular integral over the two-dimensional momentum. The integral is understood with a short-distance regulator. Using an exponential regulator,
\begin{equation}
I_a(R)=\int_0^\infty \dd q\;q^2\ee^{-aq}J_0(qR)=\frac{2a^2-R^2}{(a^2+R^2)^{5/2}},
\label{eq:IaR}
\end{equation}
and then taking \(a\to 0^+\) at fixed \(R>0\) yields the algebraic asymptote
\begin{equation}
\chiMF'(\Rb,0)
=
\frac{\mathcal C_\chi}{v_F R^3},
\label{eq:chi_realspace_final_appendix}
\end{equation}
where \(\mathcal C_\chi\) is a convention-dependent signed prefactor. The robust point is the algebraic magnitude \(|\chiMF'(\Rb,0)|\propto 1/(v_F R^3)\), which is the scaling used in the main text.

The absorptive part follows from Eq.~\eqref{eq:chi0_ret_appendix}. For \(\omega>0\), its magnitude is
\begin{equation}
|\chiMF''(\qb,\omega)|
=
A_\chi\,
\frac{q^2}{\sqrt{\omega^2-v_F^2 q^2}}\,
\Theta(\omega-v_F q),
\label{eq:chi0_imag_appendix}
\end{equation}
where \(q=|\qb|\), up to the overall retarded-sign convention. This explicitly displays the particle-hole threshold discussed in the main text.
\section{Dressed-susceptibility parameterization and robustness analysis}
\label{app:dressed}

The full interacting $\mathrm{U(1)}$ DSL is not expected to be captured exactly by the bare bubble. A practical way to include gauge-field dressing in the reduced qubit theory is therefore to parameterize the bath kernel itself. The static long-wavelength form used in the main text is
\begin{equation}
\chiphys(\qb,0)\sim Z_\chi q^{1-\eta_s},
\label{eq:chi_static_eta_appendix}
\end{equation}
which implies
\begin{equation}
\chiphys(\Rb,0)\propto \frac{1}{R^{3-\eta_s}}.
\label{eq:chi_R_eta_appendix}
\end{equation}
The resulting coherent bus law is
\begin{equation}
J_{\mathrm{eff}}(R)\propto \frac{J_{\rm local}^2}{R^{3-\eta_s}},
\label{eq:Jeff_eta_appendix}
\end{equation}
while the dynamic relaxation rate may be written schematically as
\begin{equation}
\Gamma_1\propto J_{\rm local}^2 Z_\chi\,\omega_0^{\alpha(\eta_s,\lambda)},
\label{eq:Gamma_eta_appendix}
\end{equation}
where $\alpha$ depends on the low-frequency structure of the dressed bath and on additional parameters $\lambda$ representing temperature, doping, or disorder.

For the numerical robustness runs we use a concrete dressed-$\mathrm{U(1)}$-DSL-inspired kernel,
\begin{equation}
\chiphys^R(\qb,\omega)
=
-\,Z_\chi A_\chi\,
\frac{q^{\,2-\eta_s}}
{\sqrt{v_F^2 q^2-\bigl(\omega+\ii\gamma_{\mathrm{eff}}\bigr)^2}},
\label{eq:chi_u1dsl_numeric_appendix}
\end{equation}
with
\begin{equation}
\gamma_{\mathrm{eff}}
=
\eta_{\mathrm{reg}}
+
\gamma_{\mathrm{broad}}
+
c_T\,T
+
c_\mu\,|\mu|.
\label{eq:gamma_eff_numeric_appendix}
\end{equation}
Equation~\eqref{eq:chi_u1dsl_numeric_appendix} is intended as a broadened numerical ansatz for the robustness runs, not as a strict microscopic scaling law at all momenta. When $\gamma_{\mathrm{eff}}$ is nonzero, it rounds the asymptotic low-momentum power law below the scale $q\sim \gamma_{\mathrm{eff}}/v_F$. Accordingly, Eqs.~\eqref{eq:chi_static_eta_appendix} and \eqref{eq:chi_R_eta_appendix} then represent the underlying static scaling forms recovered when this broadening is negligible in the static sector. Here $Z_\chi$ renormalizes the overall spectral weight, $\eta_s$ modifies the low-momentum power law relative to the strict mean-field benchmark, and $\gamma_{\mathrm{eff}}$ broadens the continuum edge in a way that can depend on temperature and doping. This ansatz is not intended as a complete microscopic theory of the interacting spin liquid. Its purpose is to provide a transparent numerical deformation of the benchmark bath that captures, at a phenomenological level, the type of reshaping expected once gauge-field and interaction effects are no longer ignored.

In practical phase-space maps and reduced-dynamics simulations, one can therefore propagate uncertainty in the bath sector by varying $(Z_\chi,\eta_s,\gamma_{\mathrm{eff}})$ while keeping the reduced qubit-sector derivation unchanged.

\section{Numerical implementation and observables}
\label{app:numerics}

This Appendix details the numerical workflow used to generate the main-text results. A bath kernel is first generated on finite \((q,\omega)\) grids and then converted into the static benchmark exchange \(J_{\mathrm{eff}}(R)\), the local dissipative scales \(\Gamma_1\) and \(\Gamma_\phi\), and the nonlocal cross-damping scale \(\Gamma_{AB}(R)\). In the benchmark runs, \(\Gamma_1\) denotes the local single-qubit downward relaxation scale \(\Gamma_{AA}^{\downarrow}\), while \(\Gamma_\phi\equiv \Gamma_{AA}^{\phi}\) denotes the coefficient of the local longitudinal-dephasing dissipator and \(\Gamma_{AB}\) the nonlocal cross-damping channel. With the spin normalization \(S^z=\sigma^z/2\), the contribution of this longitudinal coefficient to the physical single-qubit pure-dephasing rate is \(\Gamma_{AA}^{\phi}/2\). The resulting bus metric \(Q_{\mathrm{bus}}\) is then used to identify three representative operating sectors, labeled \emph{deep viable}, \emph{boundary}, and \emph{outside}, at which the reduced master equation is solved.

We use two complementary reduced-dynamics implementations. The first is a Lindblad implementation with fixed effective rates extracted from the susceptibility-derived kernels. The second is a Redfield-like implementation in which the bath spectral tensor is retained more explicitly. Here ``Redfield-like'' denotes an effective implementation in which the bath input is reconstructed from the susceptibility-derived spectra used throughout the manuscript, not from a microscopic calculation of the complete ordered correlator tensor of the interacting DSL. In the spin-isotropic benchmark used here, these spectra define an effective bath matrix that is diagonal in physical spin-component space and has local and nonlocal entries in qubit-position space. This two-site matrix is implemented through symmetric and antisymmetric collective channels with spectra
\begin{equation}
\mathcal S_{\pm}(\omega)
=
\mathcal S_{\rm loc}(\omega)
\pm
\mathcal S_{\rm nl}(\omega),
\end{equation}
coupled to \((S_A^a\pm S_B^a)/\sqrt{2}\) for \(a=x,y,z\). In the finite-grid implementation, positive floors and clipping of the reconstructed nonlocal spectrum keep the collective spectral weights non-negative. This construction provides a spectrally resolved robustness check within the effective susceptibility model, not a microscopic Bloch--Redfield calculation of the full interacting spin-liquid bath.

Concretely, Fig.~\ref{fig:bath_response} is obtained by evaluating the benchmark kernel of Eq.~\eqref{eq:chi0_ret} on a finite \((q,\omega)\) grid and plotting its absorptive and static components. Figure~\ref{fig:exchange_rates} uses the same susceptibility-derived input to extract \(J_{\mathrm{eff}}(R)\), \(\Gamma_1\), \(\Gamma_\phi\), and \(\Gamma_{AB}(R)\). In the numerical implementation, the finite-grid radial transform from momentum space to real space is regularized by applying a smooth cutoff function on the momentum grid before the transform, in order to suppress truncation-induced oscillations and stabilize the long-distance response relevant for the bus analysis; the qualitative trends reported below are insensitive to reasonable variations of this numerical regularization. Figure~\ref{fig:phase_map} evaluates the resulting operational metric \(\Qbus\) over the dimensionless variables \(x=\omega_0R/v_F\), \(\theta=T/\omega_0\), and \(m=\mu/\omega_0\). The phase map is constructed by computing reference exchange and dissipative curves on a discrete \(R\) grid, interpolating them to the values of \(x\) used in the scan, and applying simple temperature- and chemical-potential-dependent weighting functions rather than recomputing the full bath susceptibility at every \((\theta,m)\) point. Figure~\ref{fig:robustness} repeats the same reduced pipeline after replacing the benchmark kernel by the dressed-\(\mathrm{U}(1)\)-DSL-inspired susceptibility ansatz of Eq.~\eqref{eq:chi_u1dsl_numeric}. Finally, Fig.~\ref{fig:dynamics} solves the reduced two-qubit master equation, Eqs.~\eqref{eq:effective_master}--\eqref{eq:dissipator_sim}, at representative deep-viable, boundary, and outside operating points selected from the low-temperature, low-doping slice of Fig.~\ref{fig:phase_map}. The time-domain evolution is implemented in the computational basis
\begin{equation}
\{|00\rangle,|01\rangle,|10\rangle,|11\rangle\}.
\label{eq:comp_basis_appendixD}
\end{equation}

The principal observables used throughout the time-domain analysis are standard two-qubit diagnostics: the concurrence~\cite{Wootters1998},
\begin{equation}
C(\rho)=\max\left(0,\lambda_1-\lambda_2-\lambda_3-\lambda_4\right),
\label{eq:concurrence_appendixD}
\end{equation}
where the $\lambda_i$ are the square roots, in descending order, of the eigenvalues of
\begin{equation}
\rho(\sigma_y\otimes\sigma_y)\rho^*(\sigma_y\otimes\sigma_y),
\label{eq:rho_tilde_appendixD}
\end{equation}
the Bell-state fidelity
\begin{equation}
F_{\mathrm{Bell}}(t)=\langle\Phi_{\mathrm{target}}|\rho(t)|\Phi_{\mathrm{target}}\rangle,
\label{eq:bell_fidelity_appendixD}
\end{equation}
the purity
\begin{equation}
P(t)=\Tr\,\rho^2(t),
\label{eq:purity_appendixD}
\end{equation}
and the local polarization histories
\begin{equation}
\langle\sigma_A^z(t)\rangle,\qquad \langle\sigma_B^z(t)\rangle.
\label{eq:sz_obs_appendixD}
\end{equation}
The bath is regarded as operationally useful as a spin bus when appreciable entanglement develops on a timescale shorter than the coherence loss set by \(T_2\).
To make the robustness question explicit at the observable level, we compare the benchmark mean-field and dressed-$\mathrm{U(1)}$-DSL-inspired reduced dynamics for the same three operating sectors. The expanded concurrence, Bell-fidelity, purity, and polarization-history traces are collected in the Supplemental Material. Together they show that the dressed bath preserves the qualitative sector structure identified in the main text, while generally shifting the balance toward stronger damping and weaker Bell-state performance.

The supplemental figures provide an observable-level check of the same operating-sector structure summarized in the main text. The deep-viable sector remains the most favorable for exchange-driven entanglement, while the boundary and outside sectors show progressively weaker and shorter-lived coherence. The dressed-$\mathrm{U(1)}$-DSL-inspired kernel generally shifts the dynamics toward stronger damping, most visibly in the Bell-fidelity, Redfield-like, and polarization traces.

Thus the supplemental diagnostics support the main interpretation without adding a separate numerical narrative. The mean-field kernel gives the analytically transparent baseline, whereas the dressed-kernel comparison tests the robustness of that baseline beyond mean field. The resulting picture is a conditional many-body interconnect whose usefulness is controlled by a finite exchange-dominated window, not by a universally protected low-loss mechanism.


\bibliography{mybib}

\end{document}